# Chaos and the dynamical evolution of tidal capture binaries in globular clusters

**Rosemary A. Mardling**

Mathematics Department, Monash University, Clayton, Victoria 3168, Australia

**Globular clusters cores harbour many low-mass X-ray binaries [1], cataclysmic variables [2], and millisecond pulsar binaries [3] which are likely to have been formed via the process of tidal capture [4]. Tidal capture binaries were originally thought to be responsible for halting core collapse and for subsequent re-expansion via the process of "binary heating" [5]. The standard model [6], [7] now suggests that these binaries are no longer viable as a direct energy source. We present the results of a study which suggests that tidal capture binaries may indeed be a significant direct source of energy for the cores of globular clusters. We show that following capture, these binaries go through a short, violent chaotic phase, with the orbital eccentricity (and hence the tidal energy) suffering large changes on a very short timescale. This is followed by a long quiescent phase in which the tidal energy remains small while the orbit circularizes only via normal dissipative effects. This quiescent phase allows captured main sequence stars to evolve in the absence of large tides, a fact important for the production of low-mass X-ray binaries and cataclysmic variables.**

The extremely high stellar densities ($\sim 10^5 M_\odot$ pc$^{-3}$) and low velocity dispersions found in the cores of globular clusters make it possible for binaries to form via tidal capture [4], a process which involves conversion of orbital energy into tidal energy for stars which pass within a few stellar radii of each other. Such binaries represent less than 20% of all encounters, the rest resulting in collision or disruption at or near the time of capture (R. A. Mardling, manuscripts submitted).

Even the majority of those binaries which survive capture are thought to eventually merge. The standard model for the evolution of tidal capture binaries [6], [7] is based on a calculation done by Press and Teukolsky [8] which gives the amount of orbital energy transferred to the tides during the first periastron passage. This model assumes that the energy transferred during subsequent periastron encounters is independent of the oscillatory state of the stars and is given approximately by the Press and Teukolsky [8] formula. The orbit is supposed to then circularize on a very short timescale ($\sim 10$ yr) [6] - much shorter than the mode damping timescale ($10^4 - 10^6$ yr) [6], [7], causing the stars to expand so much that a common envelope phase is entered, with subsequent orbital decay destroying the binary and removing it as a direct heat source for the core [6]. Mass loss resulting from this process can indirectly contribute to the core dynamics as the cluster attempts to sustain virial equilibrium although, assuming the standard model, this source of core energy implies core luminosities at odds with observations [9]. The fact that globular clusters harbour in their cores almost 20% of all low-mass X-ray binaries (LMXBs) while they contain only about 0.05% of the mass of the galaxy [10], indicates that the tidal capture process is probably more efficient at producing stable binaries than the standard model implies. Other mechanisms such as neutron star exchange into primordial binaries have been put forward as viable mechanisms for their formation [11], although it is not clear that this process can produce short period LMXBs.

Binary heating is a process through which field stars can increase their kinetic energy at the expense of a binary's orbit. Because tidal capture binaries are supposed to circularize so quickly, those that do not merge essentially exist only as extremely hard binaries (their binding energy is much greater than the average kinetic energy of field stars). Interactions with such hard binaries will be rare (compared to softer binaries), and when such an event does occur, the likely outcome is ejection of all three stars from the system if not from the cluster [12], again contributing to the core energy indirectly.

The major source of binaries is now thought to be primordial binaries [13], for which there is now a variety of evidence [14]. It has been inferred that 10% of all stars are the primary of a binary [15]. Although all but the hardest of these binaries are likely to be destroyed in the core through interactions with other stars, those left will provide a significant source of core energy [16].

We present here the results of a study (R. A. Mardling, manuscripts submitted) which suggests that the tidal capture process is more efficient at producing stable binaries than the standard model implies. Previous studies of the dynamical evolution of capture binaries have been restricted to a few orbits [17], [18], [19] because of the extremely long orbital periods following capture (the timestep in any integration procedure is restricted by the dynamical timescale). This problem has been overcome by taking advantage of the fact that for a large part of a highly eccentric orbit, the orbit and oscillations are essentially decoupled, so that an analytic solution is possible for this section of the orbit. We have been able to follow the evolution of capture binaries for several thousand orbits while conserving the total energy and angular momentum of the system to at least one part in $10^4$.

The model used is based on a self-consistent normal mode analysis devised by Gingold and Monaghan [17] in which the orbit and modes of oscillation are free to exchange energy in either direction. A binary is modelled by a point mass and a non-rotating polytrope of index 1.5, representing a compact object and a fully convective low-mass main sequence star respectively. Although the model is linear in the mode amplitudes, work in progress (R. A. Mardling, in preparation) indicates that the general behaviour of interest here is not affected by the inclusion of the neglected non-linear terms, except for systems close to disruption.

The model may be started with arbitrary orbital eccentricity and periastron separation. The mass ratio and the number of modes included in the calculation are also arbitrary. We find that two distinct types of behaviour are possible for the dynamical evolution of eccentric binaries: for a range of periastron separations, eccentricities and tidal energies the orbital motion is chaotic, while for all other sets of values the motion is periodic (see Fig. 1). The type of behaviour is revealed by plotting the eccentricity against periastron passage. Fig. 2 compares two initially close chaotic orbits with a non-chaotic orbit. The chaotic orbits are capable of transferring large amounts of energy to and from the polytrope, with the energy transfer during any one periastron passage (except the first few) being up to an order of magnitude larger than the Press and Teukolsky [8] formula predicts. This reflects the fact that the energy transfer does indeed depend on the oscillatory state of the stars. In contrast, non-chaotic orbits extract very little energy. The chaotic nature of the solution has been verified a number of ways (R. A. Mardling, manuscripts submitted), including calculating the Lyapunov exponents of the system [20].

Capture orbits are started with eccentricity $e = 1$. Fig. 3 shows the evolution of a capture orbit with an initial periastron separation of 3 stellar radii (the calculation is actually started far away from periastron). The eccentricity is bounded below by $e \simeq 0.5$; at around the $2000^{\text{th}}$ orbit, the solution meets the chaos boundary appropriate to the amount of tidal energy present (see Fig. 1). The salient feature is the non-zero average eccentricity; permanent circularization may only be achieved via normal dissipative processes. Fig. 4 shows the "evolution curve" in (periastron separation-eccentricity) space which may be compared with the curve of constant orbital angular momentum, $C_J$. The difference between theses two curves indicates that a significant amount of angular momentum is transferred to the polytrope. Also shown is $C_0$ (see Fig. 1). The response of the system to dissipation is illustrated schematically by the shaded region and may be determined as follows. For a given periastron separation and eccentricity, the tidal energy will decrease and in consequence, the minimum possible eccentricity will increase (see Fig. 1). On the other hand, the maximum eccentricity attainable must decrease. Since the angular momentum transferred to the polytrope is proportional to the tidal energy present [18], it must also decrease (for a given eccentricity), resulting in the evolution curve moving inexorably towards $C_J$ until it reaches the point where $C_0$ crosses $C_J$, after which the system will no longer be chaotic. We are assuming in this argument that any change in the bulk rotation of the stars is negligible, as is any mass loss or transfer.

The chaotic phase is likely to be short because the rate at which the modes dissipate energy is proportional to the mode energy present [21], and also because mode-mode interactions may excite higher order modes which have shorter mode damping timescales (R. A. Mardling, in preparation). This phase ends when the system has dissipated an amount of energy equivalent to the binding energy of a binary with orbital parameters equal to those at the intersection of $C_0$ and $C_J$. In constrast, the modes carry little energy when the system is not chaotic, so that a long quiescent phase of circularization follows. The angular momentum transferred to the stars will be truly negligible during this phase, so that the evolution in (periastron separation-eccentricity) space follows the curve of constant angular momentum (see Fig. 4), resulting in a binary with a radius of twice the periastron separation at capture, as is generally assumed. The quiescent phase may last for up to $10^8$ years (R. A. Mardling, manuscripts submitted) making these binaries available as a direct energy source for much longer than previously thought (for comparison, a capture binary with initial periastron separation of 3 stellar radii which has circularized to the point where $e = 0.85$, will have the same binding energy as a circular binary with radius 20 stellar radii).

These results also have significant implications for the formation of LMXBs, cataclysmic variables, and millisecond pulsar binaries. Although there is likely to be some mass loss during the violent chaotic phase, the quiescent phase which follows allows captured main sequence stars to evolve in the absence of large tides. Mass transfer can then begin as such stars enter the giant phase. A study by Kochanek [18] suggests that stars with masses smaller than $0.7 M_\odot$ which are captured by a neutron star will not survive the large tides implied by the standard model. The present study instead suggests that such binaries may exist. A case in point is the binary pulsar PSR1718-19 [22], whose companion mass can be estimated at around $0.2 M_\odot$ if tidal capture is assumed (R. A. Mardling, manuscripts submitted). S. Thorsett has estimated the companion mass to be $0.14 - 0.25 M_\odot$ (R. Wijers, private communication), which would appear to vindicate the tidal capture-quiescent circularization scenario, although conclusive evidence is still needed.

It is curious that the range of periastron separations for which capture is possible in a globular cluster (up to about 3.4 stellar radii for equal mass stars of which one can be modelled with a polytrope of index 1.5, and with a relative velocity at infinity of 10 km sec$^{-1}$ [6]) corresponds so closely with the range for which the orbital motion is chaotic. Because little energy is transferred to the tides in the periodic type of solution, it is reasonable to conclude that the chaotic behaviour following capture is crucial to the formation of binary stars which are stable against ionization by nearby stars. For more centrally condensed stars, the capture cross sections in units of stellar radii reduce (except for very low velocities at infinity) [23]. Thus we might expect that binaries formed by tidal capure which are stable against mass transfer (initially, at least) tend to be formed from at least one low mass star with a convective outer layer.

This model may be tested by measuring the orbital period derivatives of globular cluster eccentric binaries containing a main sequence star. One should find a very small period derivative, in contrast to what the standard model predicts.

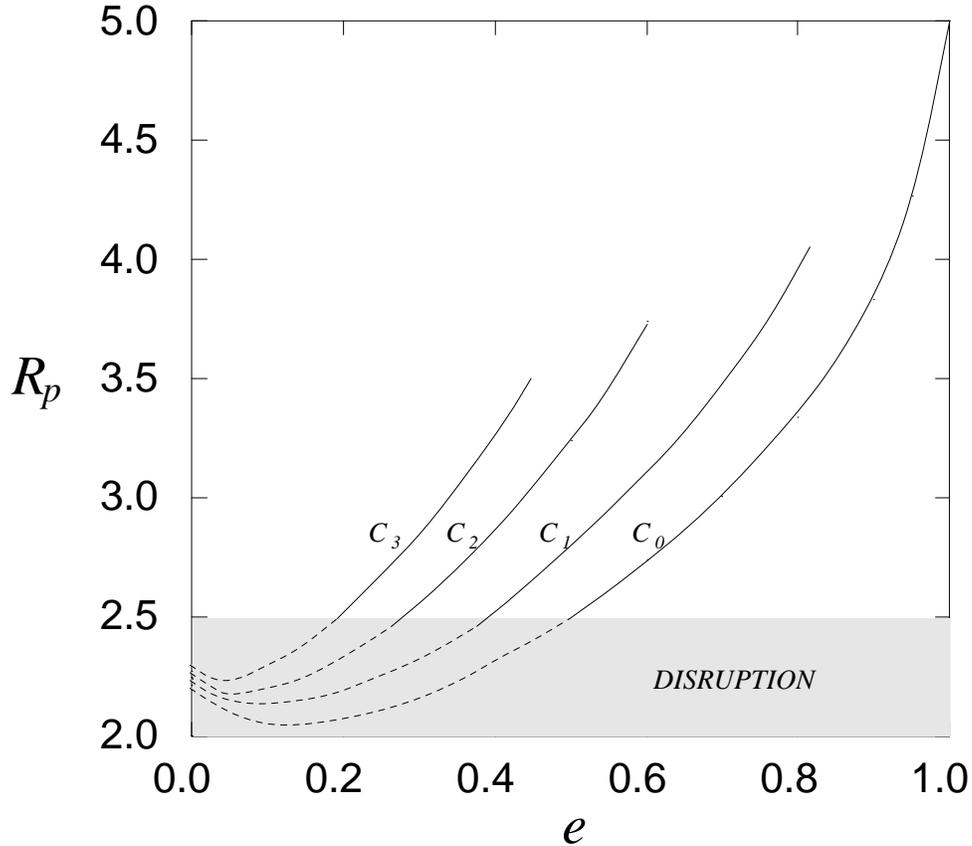

Figure 1: The chaos boundaries for equal mass stars, one of which is compact. A binary orbit with periastron separation $R_p$, eccentricity $e$ and tidal energy $x$ is chaotic if it falls below the curve $C_x$. $R_p$ is in units of a stellar radius. These curves are calculated as follows. The polytrope is started with initial tidal energy $x$ and the initial periastron separation $R_p$ and eccentricity $e$ are chosen. Another solution is calculated with the periastron separation varied slightly and the two solutions are compared. A point $(R_p, e)$ is labelled chaotic if there is clear sensitivity to initial conditions. The curves terminate at points which correspond to zero *total* energy. The Roche limit for synchronized circular binaries is around 2.5 stellar radii for equal mass stars, and we assume that this is roughly correct for systems with non-zero eccentricity. Because the model is linear in the mode amplitudes, it underestimates the Roche distance making it is possible to calculate the dashed parts of the curves.

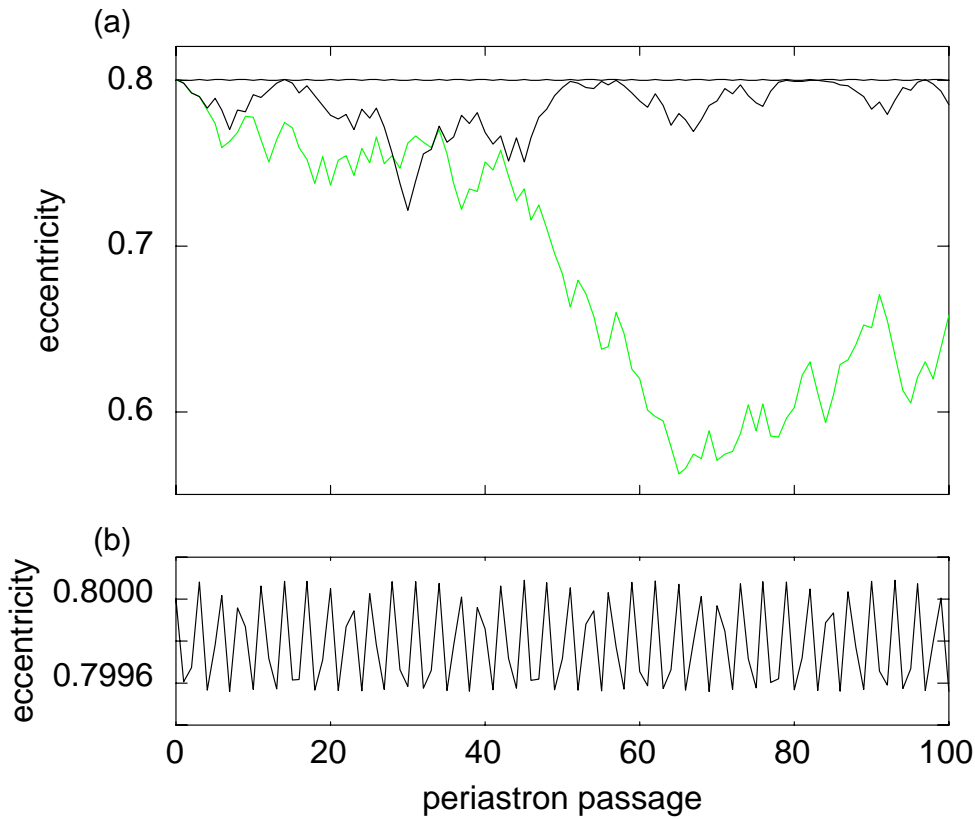

Figure 2: Chaotic versus periodic orbital motion. The uppermost curve in (a) is a periodic orbit, shown enlarged in (b). Very little energy is exchanged between the orbit and the tides, as is made apparent by the small range in eccentricity. The initial periastron separation was taken as $3.2R_*$ ($R_*$ is the radius of the polytrope). For periodic orbits, the amplitude of variation of the eccentricity is almost always of the order of the change in eccentricity after the first periastron passage. The other two curves shown represent the general behaviour exhibited by chaotic orbits. The initial periastron separations were taken to be $2.9R_*$ (grey curve) and $2.90001R_*$, a difference of about 4 parts in $10^6$. The two orbits are in synchrony until the $4^{\text{th}}$ orbit, and thereafter diverge. The total change in eccentricity for the 100 orbits shown is around 3 orders of magnitude more than for the periodic orbit, far more than is accounted by the difference in inital periastron separation. Each orbit shown in this figure was started at apastron with zero initial oscillation energy and an initial eccentricity of 0.8. The periastron separation prescribed is that which the orbit would achieve if unperturbed. The mass ratio was unity.

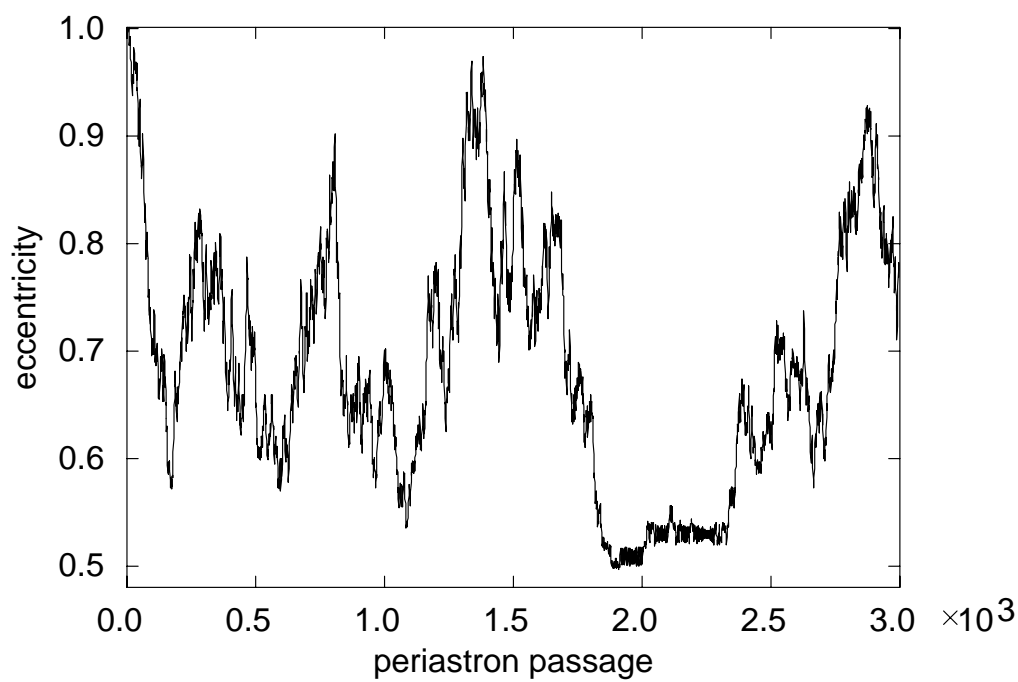

Figure 3: The evolution of a capture orbit (equal masses). This orbit was prescribed an initial periastron separation of $3R_*$ and the calculation was started at a separation of about $40R_*$. The eccentricity follows a quasi-random walk, except around the $2000^{\text{th}}$ orbit, where the solution meets the chaos boundary and becomes periodic for several hundred orbits (see text and R. A. Mardling, manuscripts submitted). The average eccentricity is non-zero; the orbit can circularize permanently only via normal dissipative processes. The $l = 2, 3$ and $4$ $f-$modes [21] were included in this calculation.

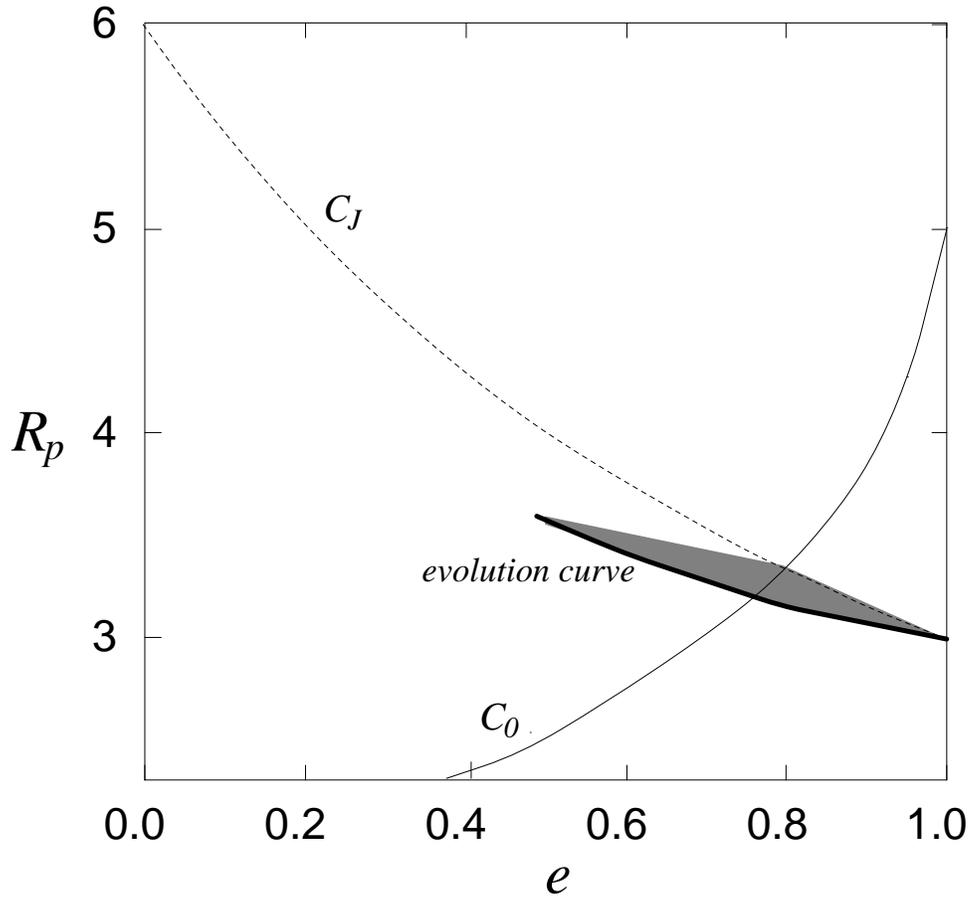

Figure 4: The evolution curve for the model shown in Fig. 3 and the effect of dissipation. The evolution curve is calculated by plotting a point at each periastron passage. These points wander up and down the "curve" (compare to Fig. 3) which actually has a width (for a given eccentricity) of about $0.02R_*$. The shaded region indicates the response of the system to dissipation (see text), and represents the short, violent chaotic phase of the capture binary's life. The evolution curve moves towards the point at which the curve of constant orbital angular momentum, $C_J$, crosses $C_0$, the chaos boundary corresponding to zero tidal energy. The system then enters a long quiescent phase during which the binary circularizes only via normal dissipative processes, following $C_J$ until it is a circular binary with a radius of twice the initial periastron separation.